# Sonifying I²S Transport Signals to Detect Transmission Faults

Stephen Roddy
*Radical Humanities Laboratory &*
*Digital Humanities Department*
*University College Cork*
Cork, Ireland
sroddy@ucc.ie

***Abstract*— This paper outlines a sonification design to support fault detection in the transmission of I²S transport signals. I²S is a protocol for communicating real-time digital audio between integrated circuits that, while in wide and general use, does not include built-in error detection. Moreover, given the nature of the protocol transmission faults affecting timing, framing and alignment can be difficult to identify using conventional visual methods. The proposed design addresses this with an approach informed by Audification, wherein oversampling controls temporal rescaling to render protocol structure (SCK and WS) and payload data (SD) across separate stereo channels. A preliminary computational feasibility study was carried out to measure feature-space separability of I²S faults in the generated auditory representations as opposed to listener performance. It evaluates the design across several payload types and error conditions including jitter, bit-slip, and word-length errors. Class separability was assessed through clustering analyses of extracted features. The evaluation results show that while oversampling produces systematic changes in feature values, it does not meaningfully improve separability between error classes. However, a modest but consistent improvement in separability is observed as a function of the joint representation of structural and payload information across channels. The findings suggest that feature-space separability in sonified communication protocol data may be dependent on the integration of complementary information streams, rather than on signal scaling alone.**

## I. Introduction

### A. Internet of Sounds

The Internet of Sounds (IoS) explores the integration of Sound and Music computing techniques with Internet of Things infrastructures to support novel sonic practices across creative, cultural, industrial and scientific domains [1]. It unites disparate interests across engineering and humanities fields with a focus on sensor networks for sensing, acquiring, processing, actuating, and exchanging sound-related data in both musical and non-musical contexts [2]. Being an interdisciplinary field grounded in the study of technology there are currently a number of open technical questions that focus on latency, jitter and synchronization across network communications layers, privacy and security both on-device and on-network, and strategies for analysing and representing complex IoS data [1], [3], [4]. While connectivity across the IoS is enabled by higher level networking protocols [1], the reliable exchange of audio data matters across the entire signal chain, both on-network and on-device. While IoS research has tended to focus on network layers, the low-level transport of audio signals between integrated circuits (ICs) within a given edge device is of critical importance to the delivery of high-quality audio.

### B. The I²S Protocol

In this regard, the I²S protocol is an essential 'last mile' physical connection for shifting audio data across ICs within a device [5], [6], [7]. I²S was developed by Philips Semiconductors in 1986 to transmit two-channel PCM audio between ICs [7]. Since then, it has become a widely used serial interface for real-time digital audio within electronic devices. However, I²S does not include built-in error detection, and so transmission errors can still occur because of timing, clocking, or signal-integrity problems. These faults can compromise the integrity of digital audio transmission. Moreover, given the high I²S transmission speeds and information dense payloads these faults can be difficult to detect using conventional visual methods. I²S, operating through synchronised clock and data lines, is susceptible to transmission faults like jitter, bit-slippage, timing misalignment and framing errors which degrade system performance. As the IoS matures and expands the volume and complexity of data passing between ICs on IoS edge devices is set to dramatically increase. As such the ability to detect subtle transport-layer errors will become increasingly important.

### C. IoS Sonification

In the IoS, existing research in sonification, the mapping of data to sound for the purposes of analysis and communication [8], has focused on the network [9], [10] and device layers [10], [11] with an expanded Sonification-Enabled IoS Network schema proposed for guiding the integration of sonification in IoS contexts [9]. However, there has been little to no research exploring low-level intra-device audio buses like I²S [1], [11], [12]. The following section outlines a design for the sonification of I²S data that is intended to support the detection of transmission faults. This design is then evaluated for a synthetic I²S dataset to determine the efficacy of the design. The sonification design described below adapts a specific sonification technique termed Audification [13]. Audification involves 'the direct translation of a data waveform into sound' [14]. It is well suited to complex signals where subtle changes are encoded in high resolution time-series data with a wave-like structure measured at constant sampling intervals (e.g. vibrational data). It is often coupled with time-scaling so that data recorded on longer timescales can be compressed for listening in shorter, more realistically manageable sessions. As such audification is an appropriate basis for the development of an I²S sonification system that converts changes in voltage over time to changes in acoustic parameters. The method explored here is an audification-based approach to sonification where direct rendering is augmented through design decisions concerning the optimal representation of multiple different data streams.

## II. I²S Sonification Design

The I²S protocol involves three distinct data lines.

$$f_{SCK} = f_s N_{bits} N_{ch} \quad (1)$$

The bit clock (SCK) provides a synchronous timing reference for the bit-level synchronisation of the serial audio data stream (SD). SCK frequency is given by the I²S sample rate ($f_s$) multiplied by the bits per sample ($N_{bits}$) and the number of channels ($N_{ch}$).

$$f_{WS} = f_s \quad (2)$$

The word select clock (WS) is synchronised to the sample rate with High/Low oscillations indicating which stereo channel a given word on the SD line belongs to. SD encodes the audio data payload. In real-world I²S systems SCK operates in the MHz range, WS at the audio sample rate (41-96 kHz) with SD clocked in relation to SCK. However, this implementation renders simulated I²S transport signals at a 48 kHz audio playback rate. The SCK runs at 24 kHz without oversampling. Human hearing operates in the range of roughly 20 Hz to 20 kHz and as such a direct rendering of the signal in audio would produce an SCK signal that is largely imperceptible to a listener. To address this the signal was temporally rescaled into the audible domain via oversampling. In the context of the sonification, this approach retains the relative timing relationships between all three signal lines and has the effect of extending individual bit transitions in time. A number of techniques are employed to create the simulated I²S signals.

$$x_{int}[n] = \text{int32}(x[n](2^{31} - 1)) \quad (3)$$

First, input audio samples x[n] are converted to 32-bit integer representations before being unpacked and flattened into the SD payload bits as normalised, bi-polar values with bit positions indexed by $m$.

$$SD[m] \in \{-1, +1\} \quad (4)$$

Oversampling is applied by repeating each *SD* bit over $O$ output samples at index $p$.

$$s_O[p] = s\left[\left\lfloor \frac{p}{O} \right\rfloor\right] \quad (5)$$

A one-clock delay is then added to the SD line to keep it in line with the I²S protocol spec. The SCK waveform is generated as a square wave at 24 kHz (without oversampling) while the synchronised WS is a square wave changing once per word at 750 Hz.

$$f_{SCK}^{son} = \frac{f_{out}}{2O} \quad (6)$$

At an $O$ of 4, an audio rate of 48 kHz ($f_{out}$) for example, the sonified SCK ($f_{SCK}^{(son)}$) is heard as a steady square wave tone at 6 kHz with associated odd-harmonic partials resulting from the square wave. For a bit-depth of 32 ($N_{bits}$) and the same oversampling factor and audio rate, the sonified WS is heard as a continuous low-frequency square wave component of 187.5 Hz with its own associated partials. For a bit-depth of 32 ($N_{bits}$) and the same oversampling factor and audio rate, the sonified WS is heard as a continuous low-frequency square wave component of 187.5 Hz with its own associated partials.

$$f_{WS}^{son} = \frac{f_{out}}{2N_{bits}O} \quad (7)$$

Both tones have a buzzy timbre characteristic of square-waves. While structural protocol information is presented on the left channel, the information-bearing SD payload is presented on the right. This is heard as a noisy wideband texture with a spectral shape that transforms on the basis of its payload. The system is written in Python with final sonification files written to 16-bit WAV output for playback and analysis. The separation of structural protocol information, (SCK and WS) from payload information (SD), and representation of these signals across left and right channels is intended to support fault detection by increasing the perceptual salience of error states.

Errors in I²S transmission are generally the result of disruptions to timing, alignment, or data framing. This includes issues like clock jitter, bit-slip or misalignment between SD and SCK and inconsistent word lengths. These errors can disrupt the integrity of both the protocol signals and the data payload (SD). The sonification design presented here aims to render these errors perceptually salient to a listener by separating structural and payload information across stereo space. Three additional fault classes: signal dropouts, discontinuities, and persistent line faults, were not considered in the design and evaluation. Dropouts and discontinuities generally produce abrupt breaks, clicks, and silences that are highly perceptually salient already, while persistent line faults can produce frozen or highly repetitive signals that are also readily detectable. The present approach focuses instead on fault classes that are less obvious and thus provide a more demanding test of the sonification design.

## III. Evaluation & Results

### A. Evaluation Setup

An evaluation was undertaken to (i) test whether oversampling improves the separability of I²S error conditions and (ii) assess whether the joint representation of structural and payload information supports separability better than the presentation of either channel in isolation. Further inferential testing was used to characterise the effect of oversampling on the extracted feature space and clustering quality across these representations. A dataset of 300 sonifications, each 2 seconds in duration, was produced for the evaluation. The sonifications were analysed across three representations: left channel (WS+SCK), right channel (SD), and a joint (stereo-derived) representation made by concatenating the left and right channel features. There were three payload types used in the I²S sonifications. These were introduced to ensure a level of heterogeneity across payload content and not treated as independent variables in the evaluation. Five variants were created for each payload type to provide a more robust dataset and reduce the impact of any one example on the evaluation. The sinusoidal variants differed in frequency, while the noise and near-silence variants used distinct patterns of random noise. Payloads were chosen to represent a range of signal complexity levels, from analytically simple to highly stochastic. Near-silence payloads minimise the influence of the underlying signal, allowing transport-layer information to dominate more of the sonification. Sinusoidal signals reveal spectral distortions, and noise payloads provide a stress-test of error detection under conditions of high spectral complexity. The dataset also contained four error conditions. These included a clean (error-free) condition, a bit-slip condition involving misalignments between the SD and the SCK, a word length error condition with mismatch between expected word lengths, and finally a jitter condition where timing on the SCK becomes irregular. Each condition models a different class of $I^2S$ transmission error affecting timing, alignment, and data framing. These sonifications were generated across five oversampling levels (2, 4, 6, 8, and 10). Oversampling level provides the primary independent variable for the analyses alongside comparisons across representations and error conditions. For the repeated-measures analyses, the experimental unit was defined by each combination of error condition, payload-type, and payload variant. Each unit was evaluated across all five oversampling levels, with oversampling providing the repeated-measures factor. This computational analysis provides a preliminary feasibility assessment rather than a measure of perceptual discriminability or listener performance. It is intended to establish whether the generated auditory representations contain sufficient feature-space structure to justify perceptual evaluation through listener testing.

### *B. Data Analysis Pipeline*

Data analysis was handled in Python using NumPy, SciPy, scikit-learn, statsmodels, and spafe. The pipeline used is largely parallel with both clustering metrics and inferential statistical analyses computed on feature vectors extracted from the audio dataset. Some additional inferential analyses were applied to sample-level silhouette scores derived from clustering. Clustering separability was used to characterise structures in the audio feature space that correspond to $I^2S$ error conditions [15]. Accordingly, clustering metrics were used to evaluate class separability across oversampling levels and channel configurations, while inferential statistical analyses were used to characterise feature-level responses to oversampling. The analysis pipeline is not intended for use in an automated $I^2S$ fault-detection system nor is the claim made that audio improves machine fault detection over analysis of the original signals.

### *C. Feature Extraction*

A set of relevant audio features across time, frequency and cepstral domains were extracted from the sonified $I^2S$ signals. These were chosen to capture both the structure of the protocol and the characteristics of the transmitted payload data. The feature set included spectral centroid and spectral flux [16], autocorrelation peak height and lag [17] and GFCC-based descriptors of energy and variability [18]. GFCCs are 'cepstral' domain descriptors increasingly used for audio signal processing tasks. The summary GFCC energy and variability features were used in the analyses. Features were independently extracted from the left and right channels. Joint representations were then created by concatenating the feature vectors for left and right channels. Spectral centroid and spectral flux were computed in frames of 2048 samples with a 512 sample hop size. GFCC descriptors were computed using spafe's [19] default 25-ms Hamming-windowed frames with a 10-ms hop, corresponding to a frame size of 1200 samples and a hop size of 480 samples at the 48-kHz sample rate. A 2048-point FFT was also selected. Features were subsequently aggregated into fixed-length feature vectors and standardised (z-score normalisation) for analysis. Global features (autocorrelation peak height and lag) were computed across the entire signal. While frame level features were selected to identify short-term spectral transformations and transitions, the global autocorrelation features were selected to identify repetitive and periodic structural patterns. The subsequent clustering, dimensionality reduction (PCA), and inferential statistical analyses were computed on these feature vectors.

### *D. Clustering Metrics Results*

Clustering analyses were used to assess class separability across oversampling levels (2, 4, 6, 8, 10) for the four error conditions: clean, jitter, bit-slip, and wrong word length. Separation ratio (derived from mean within-class and between-class distances) and silhouette score (based on pairwise distances) were computed for the left, right, and combined representations. For the left channel, clustering structure remained unchanged across all oversampling levels. While jitter was consistently separated, the clean, bit-slip, and wrong-word-length conditions were effectively coincident. This is congruent with the design which presents protocol structure (WS + SCK) on the left channel but bit-slip and framing errors primarily affect the SD information presented on the right channel. As such, the mean within-class distance was 0.0 and the between-class distance remained constant at 2.582, producing a silhouette score of 0.25 that remained constant across oversampling levels. The separation ratio ($\approx 2.58 \times 10^{12}$) is not informative as it results from division by the 0.0 value for within-class distance.Weak clustering is apparent on the right-channel where within- and between-class distances have similar magnitude. For example 2.65 vs. 2.64 at an oversampling level of 2 and 2.37 vs. 2.54 at a level of 10. This results in separation ratios close to 1 (0.99–1.07) and negative silhouette scores (−0.079 to −0.021) which are indicative of weak class separability. A slight improvement in clustering is observed with increasing degrees of oversampling. This is consistent with the sonification design which places the SD payload in the right channel resulting in a comparatively complex and variable signal.Clearer clustering is apparent in the combined representation which contains both the left and right channels. The magnitudes of between-class distances are consistently larger than within-class distances (e.g., 4.31 vs. 2.65 at OS=2 and 4.19 vs. 2.37 at OS=10). At increasing oversampling factors separation ratios (1.63–1.77) and silhouette scores (0.090–0.127) increase. While global silhouette scores provide an overall measure of clustering quality, per-sample silhouette scores were used to examine class-level behaviour in more detail. Per-sample silhouette scores suggested that jitter achieved perfect separation (mean silhouette = 1.0) in the left channel. All other classes scored 0.0 suggesting no real separation between them. Only jitter achieved positive separation (mean = 0.116) in the right channel with all other classes being negative, indicating poor clustering.

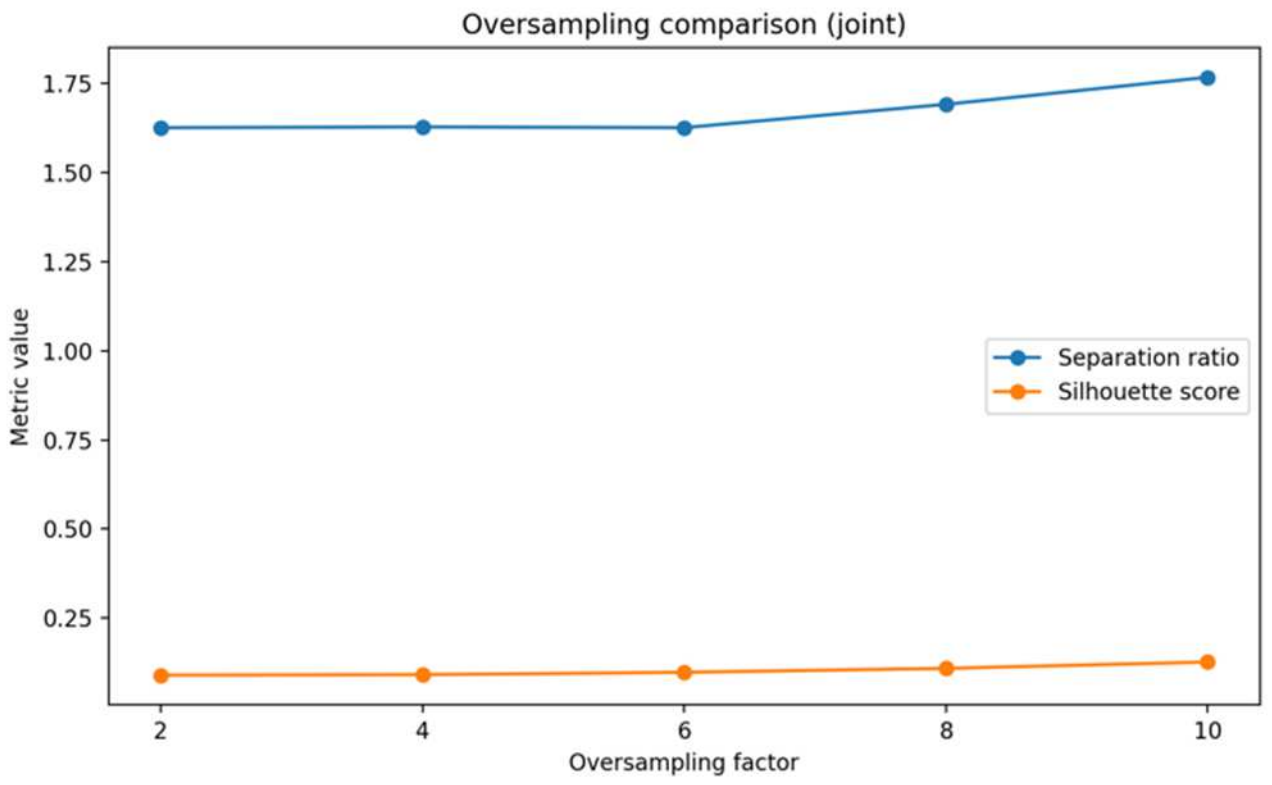

Fig. 1. Separation ratio and silhouette score for the joint representation across all oversampling levels. Higher values indicate greater feature-space separation across all conditions.

Jitter showed strong separation again in the joint representation (mean = 0.619) with the remaining classes being poorly separated (≈ -0.07). The separation results for joint representation do not represent uniform improvements in discrimination across all fault classes but are instead driven by the separation of jitter. There is still substantial overlap between the clean, bit-slip and wrong-word-length conditions.

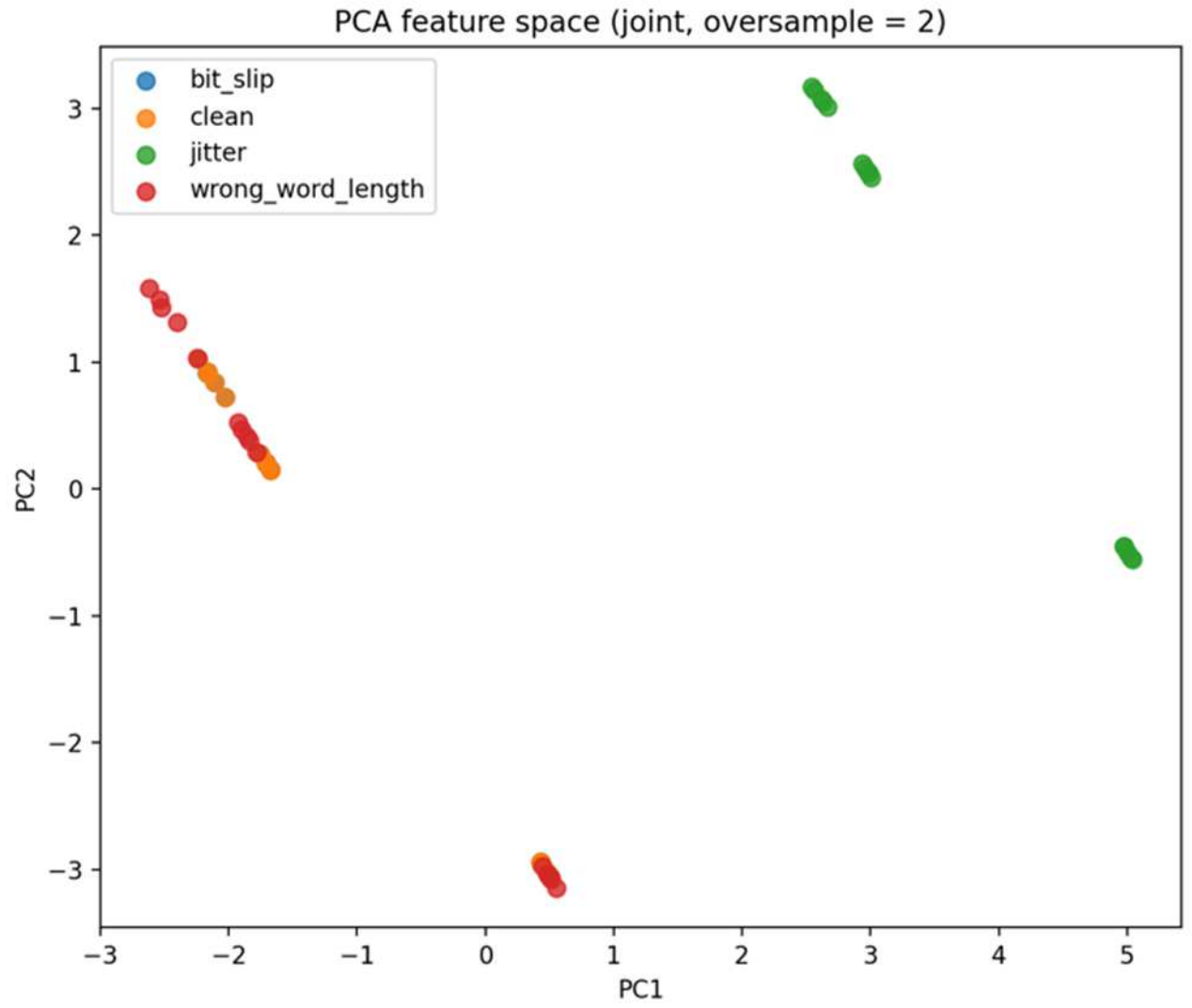

Fig. 2. PCA plot of the joint representation at OS = 2. Jitter is clearly separated, but there is substantial overlap across the remaining conditions.

Overall the results suggest that (i) oversampling has little meaningful impact on separability in the left-channel and limited impact in the right channel, but (ii) produces modest and consistent improvements in the joint representation. This suggests that greater separability is achieved with the joint presentation of sonified structural and payload information across stereo channels.

## E. *Principal Component Analysis Results*

PCA was used to support the clustering evaluation and provide a visual overview of the projected class structure across left, right and joint representations. For the left-channel PCA jitter was consistently separated across oversampling levels while clean, bit-slip, and wrong-word-length conditions largely overlapped. The relative relationships between classes in the PCA plots remained stable across oversampling levels. In the right-channel PCA, substantial overlap was observed between classes, and class boundaries remained poorly defined. This remained true across all oversampling levels, albeit with slightly tighter point distributions and modest improvements in visual separation at higher levels.

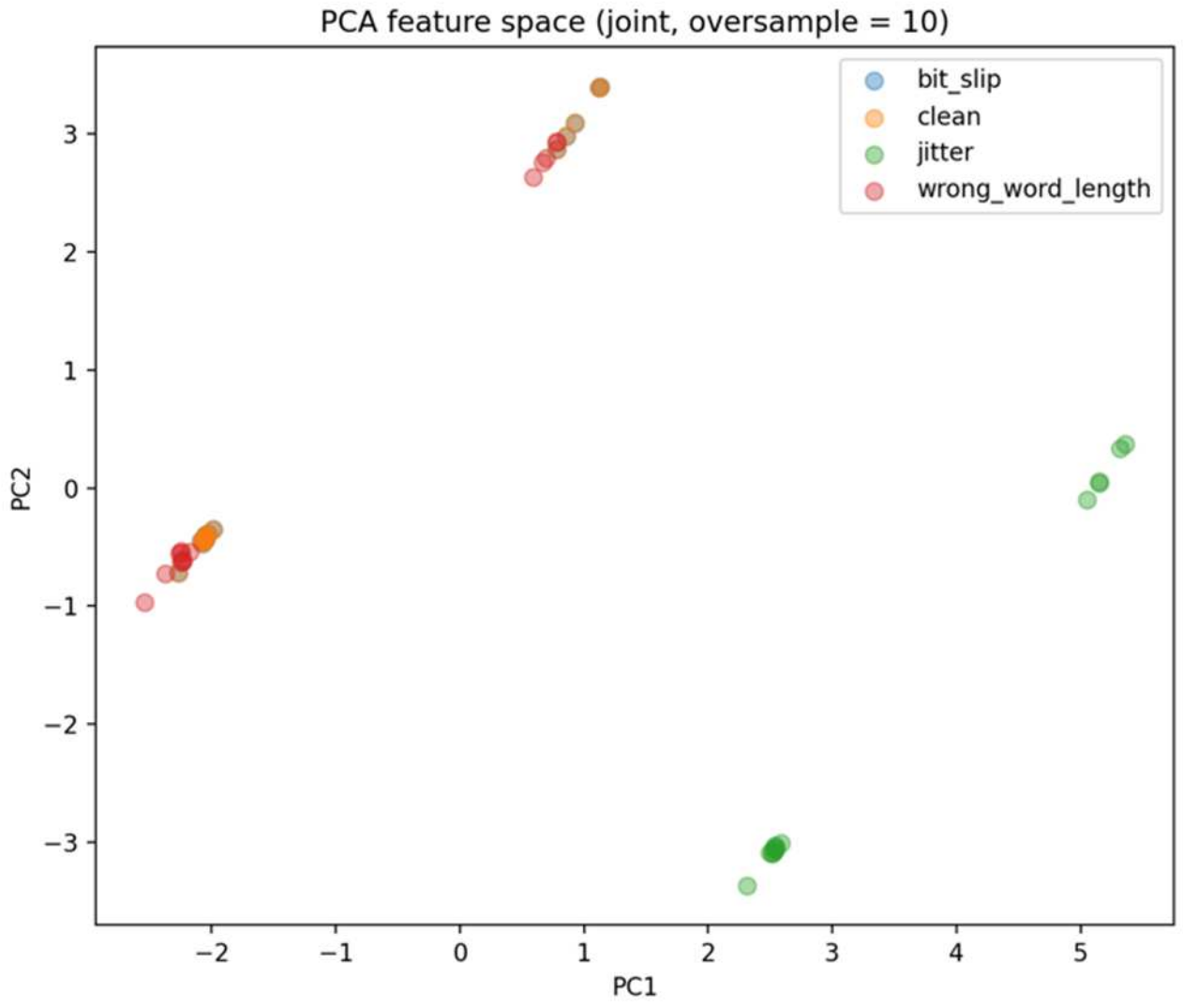

Fig. 3. PCA plot of the joint representation at OS = 10. Jitter is still well separated, but the overlap persists across the remaining conditions.

Partial separation was apparent in the joint representation with the jitter class isolated across all oversampling levels. However, there is substantial overlap between the clean, bit-slip, and wrong-word-length classes (as per figs 2 and 3). While increased oversampling produced tighter distributions, considerable inter-class ambiguity still remained. These results further indicate that greater feature-space separability is achieved through the joint representation of protocol structure, as encoded in the left channel of the sonification system, and payload information, as encoded in the right channel. Overall, the PCA supports the clustering results suggesting (i) little meaningful effect of oversampling on separability in the left channel, limited effect in the right and (ii) modest but consistent effect in the joint representation. This is largely consistent with the cross-channel separation between protocol (WS and SCK) and payload (SD) information in the original sonification design.

## F. *Repeated-Measures ANOVA & Friedman Test Results*

Repeated-measures ANOVAs and Friedman tests were used to provide further inferential support for the results of the clustering analysis by identifying significant changes in feature values and clustering quality as oversampling increased. Repeated-measures tests could not be computed for variables that remained constant across oversampling levels, as these methods require variation across conditions. Sample silhouette scores were also analysed to assess changes in clustering performance. The ANOVAs revealed that oversampling in the left-channel had a significant effect on spectral centroid ($F(4, 236)=6004.84$, $p<0.001$, $\eta_p^2=0.990$), ACF peak lag ($F=1.19\times10^{31}$, $p<0.001$, $\eta_p^2=1.000$), and GFCC energy ($F=26442.44$, $p<0.001$, $\eta_p^2=0.998$). There was a moderate effect for spectral flux ($F=19.11$, $p<0.001$, $\eta_p^2=0.245$) and a small effect on GFCC variability ($F=3.60$, $p=0.007$, $\eta_p^2=0.057$). ACF peak height was not significant ($F=0.99$, $p=0.413$). Some of these large effects, especially those for ACF peak lag, reflect the temporal rescaling as oversampling increases and are not indicative of better class separability. Friedman tests showed consistent rank ordering for spectral centroid, ACF peak lag, and GFCC energy ($\chi^2=240.0$, $p<0.001$, $W=1.0$). Moderate and smaller effects were also found for GFCC variability ($\chi^2=87$, $p<0.001$, $W=0.36$) and spectral flux ($\chi^2=60$, $p<0.001$, $W=0.25$) respectively. On the right, there was a significant effect of oversampling on spectral centroid ($F=765.64$, $p<0.001$, $\eta_p^2=0.928$), spectral flux ($F=430.49$, $p<0.001$, $\eta_p^2=0.879$), ACF peak height ($F=182.08$, $p<0.001$, $\eta_p^2=0.755$), and GFCC energy ($F=512.03$, $p<0.001$, $\eta_p^2=0.897$). It also had a significant effect on sample silhouette ($F=21.79$, $p<0.001$, $\eta_p^2=0.270$), which suggests a modest improvement in clustering quality. Friedman tests again supported these results with perfect rank consistency for spectral centroid, ACF peak height, GFCC energy and spectral flux ($\chi^2=240.0$, $p<0.001$, $W=1.0$). Smaller effects were found for sample silhouette ($\chi^2=30.73$, $p<0.001$, $W=0.128$) and GFCC variability ($\chi^2=29.71$, $p<0.001$, $W=0.124$). Feature-level effects in the joint representation were consistent with the individual channels. Sample silhouette showed a significant effect of oversampling for both ANOVA ($F=13.69$, $p< 0.001$, $\eta_p^2= 0.188$) and Friedman ($\chi^2=15.29$, $p<0.01$, $W=0.064$), supporting the modest clustering quality improvements found in the clustering analysis. The results of Friedman tests for joint representations are consistent with those of the individual channels from which they are concatenated, with the exception of sample silhouette which was evaluated on the concatenated feature space. Overall, the inferential results provide further support for the clustering analysis. While oversampling produces systematic changes in feature space, these coincide with only modest improvements to clustering quality with the most obvious gains observed for joint representations. This suggests the sonification design effectively presents complementary information across the stereo channels, and that greater feature-space separability is achieved with joint representations of these sonified data streams which represent concurrent structural and payload data.

### *G. Trend Analysis Results*

A trend analysis was carried out to provide further support for the clustering analysis by quantifying the direction, magnitude, and consistency of changes in feature values, and clustering quality (sample silhouette), as oversampling increased in left, right and joint representations.

Strong linear trends were found for spectral centroid ($\beta = -279.46$, $R^2 = 0.918$, $p < 0.001$) and GFCC energy ($\beta = -0.009$, $R^2 = 0.903$, $p < 0.001$) across the left representation. A linear deterministic relationship was found between ACF peak lag and oversampling ($R^2 = 1.0$) as expected since oversampling stretches the sonification signal in time. Despite these strong feature-level trends, the underlying clustering metrics showed no corresponding trend.

Multiple features exhibited strong trends in the right-channel including spectral flux ($R^2 = 0.966$, $p < 0.001$) and GFCC energy ($R^2 = 0.920$, $p < 0.001$). Similarly, sample silhouette showed a significant increase with a moderate effect size ($\beta = 0.0070$, $R^2 = 0.334$, $p < 0.001$). This nonetheless suggests modest improvements to clustering quality. Sample silhouette also increased significantly in the joint representation ($\beta = 0.0050$, $R^2 = 0.330$, $p < 0.001$), indicating modest but consistent improvements to clustering with oversampling. Sample silhouette could not be evaluated for the left channel and ACF peak lag did not vary and could not be evaluated for the right. Overall, these results suggest that oversampling produces systematic changes in feature values without substantially improving class separability. Modest positive trends were found in the right and joint representations. This supports the clustering results and further suggests that the separability of $I^2S$ transmission errors depends less on oversampling factor and more on the joint representation of structural and payload information.

## IV. Discussion

The results show that oversampling does not meaningfully improve class separability although the joint presentation of structural and payload information produces modest but consistent improvements in separability between error classes. However, this separation is largely driven by jitter with clean, bit-slip and wrong-word-length conditions overlapping substantially. Additional context for this result is provided by the feature level analyses where oversampling results in systematic changes in several descriptors, including spectral centroid, spectral flux, ACF measures, and GFCC energy and variability. Detectable changes on these metrics do not correspond with changes in separability.

The left channel contains our protocol structure (WS and SCK) which, being comprised of two periodic fixed frequency square waves at different frequencies, is low in entropy and highly deterministic in structure. The results of the clustering analysis and PCA reflect this, with the jitter condition being consistently isolated while the clean, bit-slip, and wrong-word-length conditions are poorly separated. This class relationship remains invariant across oversampling levels because oversampling does not change the underlying temporal structure of the two clock signals. The picture is more nuanced for the right channel which contains the SD payload of the encoded audio content, which, when sonified, becomes a noisy, high entropy, wideband signal which separates very poorly in feature space with poor results for clustering and PCA. Oversampling has limited impact on separability because, while it changes spectral centroid, flux and GFCCs it does not reveal any additional class-level structure in the signal. It simply rescales it so that features are transformed but class relationships are more or less maintained.

In the joint representation, clustering modestly improves and separability begins to emerge over increased oversampling levels. This is because the $I^2S$ error conditions tested impact the relationships between signals. Jitter primarily affects the timing of SCK, but can also impact WS (clocked to SCK) and thus lead to the incorrect sampling of SD. Word-length errors impact the alignment between WS cycles and SD data distorting word boundaries resulting in malformed data. Bit-slip directly impacts the alignment between SD and WS and SCK. The joint representation, where structural and payload data are presented along distinct channels supports improved but limited separability between error states.

Overall these results suggest that the feature-space separability of sonified communication protocol data may depend on the structured representation and integration of complementary information streams that preserve the relationships between protocol signals. This approach is intended for debugging contexts where it can provide a complementary diagnostic tool alongside traditional visual information provided by oscilloscopes, logic analysers and protocol decoders.

## V. Limitations

All the I²S signals used here were synthetic in nature. They were simulated and not captured from real-world hardware. As such the full range of signal variability and noise encountered in physical systems may not be accurately represented. Further, the evaluation examined feature-space clustering only and did not capture human performance. Nor did it draw any comparisons with outside baselines or reference benchmarks. Real-world listener studies are needed to provide this important insight [20] and complement the objective measure of separability provided by clustering. Finally, this evaluation focused on three specific error conditions (jitter, bit-slip, and word-length errors). Fault classes like signal dropouts or line faults were omitted. Future work might use real-world data to examine additional fault classes through listener evaluations.

## VI. Conclusions

This paper presented a design for the sonification of I²S transport signals and a preliminary computational feasibility study. The sonification design is intended to support the detection of faults in I²S transmission lines by listening. The evaluation showed that while oversampling produces systematic changes in feature values of sonified I²S signals, it does not meaningfully improve separability between classes of transmission error. However, the results also showed that the strongest separation between error classes was achieved through the joint representation of structural and payload information across separate channels. This result was primarily driven by separation in the jitter condition with the clean, bit-slip and wrong-word-length conditions retaining substantial overlap. This approach concurrently represents both structural and information-bearing signals across error states. These findings suggest that greater feature-space separability is achieved through the joint representation of complementary signal components as opposed to representing them in isolation, justifying further examination through perceptual evaluation and listener testing. Sonification source code, analysis code and the synthetic dataset are all made available at [21].

## Acknowledgements

OpenAI's ChatGPT was employed during the development of components of the Python code providing assistance and debugging for the implementation and analysis scripts described in Sections II and III. Generated material has been reviewed and validated by the author.